\documentstyle[12pt]{article}
\begin{document}
\newcommand{\beq}{\begin{equation}}
\newcommand{\eeq}{\end{equation}}
\newcommand{\beqa}{\begin{eqnarray}}
\newcommand{\eeqa}{\end{eqnarray}}
\newcommand{\barr}{\begin{array}}
\newcommand{\earr}{\end{array}}
\newcommand{\nonum}{\nonumber}
\thispagestyle{empty}
\begin{center}
\LARGE \tt \bf{Analogue non-Riemannian black holes in vortical moving plasmas}
\end{center}
\vspace{2.5cm}
\begin{center} {\large L.C. Garcia de Andrade\footnote{Departamento de F\'{\i}sica Teorica - Universidade do Estado do Rio de Janeiro-UERJ.

Rua S\~{a}o Fco. Xavier 524, Rio de Janeiro, RJ

Maracan\~{a}, CEP:20550-003 , Brasil.

E-mail.: garcia@dft.if.uerj.br}}
\end{center}
\vspace{1.0cm}
\begin{abstract}
Analogue black holes in non-Riemannian effective spacetime of moving vortical plasmas described by moving magnetohydrodynamic (MHD) flows. This example is an extension of acoustic torsion recently introduced in the literature (Garcia de Andrade,PRD(2004),7,64004), where now the presence of artificial black holes in moving plasmas is obtained by the presence of an horizon in the non-Riemannian spacetime. Hawking radiation is computed in terms of the background magnetic field and the magnetic permeability. The metric is singular although Cartan analogue torsion is not necessarily singular. The effective Lorentz invariance is shown to be broken due to the presence of effective torsion in strong analogy with the Riemann-Cartan gravitational case presented recently by Kostelecky (PRD 69,2004,105009). 
\end{abstract}
\vspace{0.5cm}
\vspace{2.0cm}
\newpage
\pagestyle{myheadings}
\markright{\underline{Analogue non-Riemannian black holes}}

\section{Introduction}
Several examples of analogue black holes have appeared recently in the literature \cite{1}. They are of various types and are formed in distinct material physical media such as in optical black hole media discovered by Leonhardt and Pwinicki \cite{2} in low temperature gases where light is strongly slow at velocities as low as $17 ms^{-1}$ , and also exists in dielectric moving media as recently discussed by Schuetzhold et al \cite{3} and De Lorenci et al \cite{4}. In their model two distinct optical metrics are found with consequent birefrigence and two respective horizons and optical black holes. Also the Hawking temperature \cite{5} is also found in terms of the electric field. Earlier we have shown \cite{6} that a non-Riemannian spacetime structure called acoustic torsion \cite{7} can be useful to explain certain examples of vortex acoustics in Euler or Navier-Stokes vortical viscous fluids \cite{8}. In this report we show that non-Riemannian effective black holes can be found in moving plasmas \cite{9} from the wave equation in (MHD) flows with vorticity. Here although bi-metricity is not present, since we have just one metric representing the effective black hole, birefrigence is present which is also an example of Lorentz symmetry breaking \cite{10} as recently showed by Kostelecky \cite{11} in the case of relativistic gravitational Riemann-Cartan spacetime. 
The MHD wave equation is perturbed and the perturbed wave equation is compared with the wave equation in Riemann-Cartan spacetime which breaks unitarity. The perturbation is potential although the background fluid is endowed with vorticity. Although in principle potential velocity perturbations can be contaminated by vortical background flows, this is a technical problem by specialists in vortex acoustics in water waves that in principle can be solved \cite{12}. Actually we have two metrics the first  is the Minkowski metric which is the laboratory metric while the other is the Riemannian effective metric of the background fluid endowed with the effective torsion. It is important to stress that the torsion consider here is not the torsion of helical flows so common in plasma physics \cite{13}. besides the perturbed metric. Therefore two analogue special relativistic metrics appears in the system. The metrics are not called acoustic here neither the black holes are called sonic because in fact Alfven waves decoupled from the acoustic waves in the plasma medium. The magnetic background field is chosen to be constant \cite{9}. This paper is organised as follows: In section 2 we obtain the wave equation for vortical MHD flows and how the effective metric forming whose horizon gives rise to the  non-Riemannian artificial black hole in moving plasmas.In section 3 we compute the Hawking radiation and the effective Lorentz symmetry violation  of the MHD flow. In section 4 the conclusions and discussions are presented.
\section{Non-Riemannian geometry of MHD analogue black  hole}     
In this section we shall consider the MHD field equations of a non-dissipative perfect magnetohydrodynamical gas 
\begin{equation}
\vec{E}+\frac{1}{c}\vec{v}{\times}\vec{B}=\vec{0}
\label{1}
\end{equation}
\begin{equation} 
\frac{{\partial}\vec{B}}{{\partial}t}= {\nabla}{\times}(\vec{v}{\times}\vec{B})
\label{2}
\end{equation}
\begin{equation}
{\nabla}.\vec{B}=0
\label{3}
\end{equation}
\begin{equation}
\vec{J}=c{\nabla}{\times}\vec{H}
\label{4}
\end{equation}
\begin{equation}
{\nabla}.\vec{J}=0
\label{5}
\end{equation}
\begin{equation}
\frac{{\partial}{\rho}}{{\partial}t}+{\nabla}.({\rho}\vec{v})=0
\label{6}
\end{equation}
\begin{equation}
{\rho}[\frac{\partial}{{\partial}t}{\vec{v}}-{\nabla}(\frac{1}{2}v^{2})-\vec{v}{\times}\vec{\Omega}]= -{\nabla}({\pi}+\frac{1}{2}{\mu}_{1}H^{2})-{\rho}{\nabla}g+{\mu}(\vec{H}.{\nabla})\vec{H}
\label{7}
\end{equation} 
\begin{equation}
\dot{\eta}=0 
\label{8}
\end{equation} 
where $\vec{E}$, $\vec{B}$ are the electric and magnetic induction  fields respectively while , $\vec{H}$ is the  magnetic field and ${\mu}^{*}:= 2{\mu}-{\mu}^{2}$ ,$\vec{J}$ is the electric current, connected by the relation $\vec{B}={\mu}\vec{H}$ where ${\mu}$ is the magnetic permeability and finally ${\rho}$ is the fluid density. Another constraint we use in the MHD fluid is to make the term $(\vec{H}.{\nabla})\vec{H}=0$. Here $\vec{\Omega}={\nabla}{\times}\vec{v}$ is the vorticity of MHD flow. In the paper ${\pi}$ represents the pressure with in the polytropic gas \cite{9} we have 
\begin{equation} 
{\pi}=A({\eta}){\rho}^{\gamma}
\label{9}
\end{equation}
${\gamma}$ represents the adiabatic exponent of the polytropic gas. In the approximation considered here we can consider that $A({\eta})$ is approximately equal to one since this is given by \cite{9}
\begin{equation} 
A({\eta})=(\frac{{\pi}_{0}}{{\rho}_{0}{\gamma}})exp[\frac{({\eta}-{\eta}_{0})}{c_{v}}]
\label{10}
\end{equation}
$c_{v}$ is the specific heat at constant volume. The magnetic diffusity ${\eta}^{B}$ vanishes and the magnetic Reynolds $R^{B}$ number is infinity accordingly with the expression
\begin{equation} 
R^{B}= \frac{|{\nabla}{\times}(\vec{v}{\times}\vec{B})|}{{\eta}^{B}|{\nabla}^{2}\vec{B}|}
\label{11}
\end{equation}
The MHD flows considered here are nonrelativistic or $\frac{|\vec{v}|}{c}<<1$. The next step is to linearize these MHD equations accordingly with the perturbation expressions
\begin{equation}
\vec{\Omega}=\vec{\Omega}_{0}+\vec{\Omega}_{1}
\label{12}
\end{equation}
\begin{equation}
\vec{H}=\vec{H}_{0}+\vec{h}
\label{13}
\end{equation} 
\begin{equation}
{\pi}={\pi}_{0}+{\pi}_{1}
\label{14}
\end{equation}
\begin{equation}
{\rho}={\rho}_{0}+{\rho}_{1}
\label{15}
\end{equation}
\begin{equation}
{\vec{v}}={\vec{v}}_{0}+{\vec{v}}_{1}
\label{16}
\end{equation}
where ${\rho}_{1}<<{\rho}$, $|\vec{H}_{1}|<<|\vec{H}_{0}|$ and $|\vec{v}_{1}|<<|\vec{v}_{0}|$ carachterize the perturbations. Here the quantity , $\vec{v}=\vec{v_{0}}+ \vec{v}_{1}$, ${\rho}_{0}=constant$, ${\pi}_{0}$ and $\vec{H}_{0}=constant$ represent a steady-state uniform solution of the MHD equations. The constant density of the background flow by the conservation equation in first approximation
\begin{equation}
\frac{{\partial}{\rho}_{1}}{{\partial}t}+{\nabla}.({\rho}_{0}\vec{v}_{1})+{\nabla}.({\rho}_{1}\vec{v}_{0})=0
\label{17}
\end{equation}
implies that even by assuming that the background MHD flow is imcompressible or ${\nabla}.\vec{v_{0}}=0$ the perturbed flow does not need to be unless  ${\rho}_{1}=constant$. Thus perturbing this solution we obtain the following equations
\begin{equation}
\frac{{\partial}{\rho}_{1}}{{\partial}t}+{\nabla}.({\rho}_{0}\nabla{\Psi}_{1})+{\nabla}.({\rho}_{1}\nabla{\Psi}_{0})=0
\label{18}
\end{equation}
\begin{equation}
{\rho}_{0}\frac{\partial}{{\partial}t}{\vec{v}_{1}
}+{\rho}_{0}\nabla[\vec{v}_{0}.\vec{v}_{1}]-\vec{v}_{1}{\times}\vec{\Omega}_{0}= -{\nabla}({\pi}_{1}+\frac{1}{2}{\mu}_{1}\vec{h}.\vec{H}_{0})
\label{19}
\end{equation} 
\begin{equation}
\frac{{\partial}\vec{h}}{{\partial}t}=(\vec{H}_{0}.{\nabla})\vec{v}_{1}-\vec{H}_{0}({\nabla}.\vec{v}_{1})+(\vec{h}.{\nabla})\vec{v}_{0}-\vec{h}({\nabla}.\vec{v}_{0})
\label{20}
\end{equation} 
We also consider here that the background flow is stationary or $\frac{\partial}{{\partial}t}\vec{v}_{0}=0$. A irrotational perturbation where $\vec{v}_{1}={\nabla}{\psi}_{1}$ is assumed. By considering the expression
\begin{equation}
{\nabla}{\alpha}:= \vec{v}_{1}{\times}\vec{\Omega}_{0}
\label{21}
\end{equation}
performing the line integration of this gradient yields
\begin{equation}
{\alpha}=\int{{\nabla}{\alpha}.d\vec{r}}:= \int{d\vec{r}.\nabla{\psi}_{1}{\times}\vec{\Omega}_{0}}
\label{22}
\end{equation}
where $d\vec{r}=d\vec{r}_{0}+d\vec{r}_{1}$, where $\vec{v}_{0}=\frac{d}{dt}\vec{r}_{0}$. Equation (\ref{22}) to first approximation yields
\begin{equation}
\frac{\partial}{{\partial}t}{\alpha}= {\nabla{\psi}_{1}}.\vec{v}_{0}{\times}\vec{\Omega}_{0}
\label{23}
\end{equation}
From this expression and equation (\ref{19}) we obtain
\begin{equation}
{\rho}_{0}\frac{{\partial}}{{\partial}t}{\psi}_{1}+{\rho}_{0}\vec{v}_{0}.{\nabla}{\psi}_{1}-{\nabla}{\psi}_{1}{\times}{\nabla}{\times}\vec{v}_{0}= -({\pi}_{1}+\frac{1}{2}{\mu}_{1}\vec{h}.\vec{H}_{0})
\label{24}
\end{equation}
The time derivative again of equation (\ref{19}) reads
\begin{equation}
{\rho}_{0}\frac{{\partial}^{2}}{{\partial}t^{2}}{\psi}_{1}+{\rho}_{0}\vec{v}_{0}.{\nabla}\frac{\partial}{{\partial}t}{\psi}_{1}-{\rho}_{0}\frac{\partial}{{\partial}t}{\alpha}
= -({\gamma}{{\rho}_{1}}^{{\gamma}-1}\frac{{\partial}}{{\partial}t}{{\rho}_{1}}+\frac{1}{2}{\mu}_{1}[\frac{\partial}{{\partial}t}
\vec{h}].\vec{H}_{0})
\label{25}
\end{equation} 
where we have used the simplifications above to constraint the fluid and simplify computations. Note that the first term on the RHS of the equation (\ref{25}) must be dropped because is not a first order term in ${\rho}_{1}$. Substitution of equation (\ref{20}) and (\ref{23}) into (\ref{25}) yields
\begin{equation}
(\frac{{\partial}}{{\partial}t}+{\nabla}.\vec{v_{0}})\frac{{\rho}_{0}}{{c_{A}}^{2}}(\frac{{\partial}}{{\partial}t}+{\vec{v_{0}}}.{\nabla}){\psi}_{1}-{\nabla}.({\rho}_{0}{\nabla}{\psi}_{1})+\vec{K}.{\nabla}{\psi}_{1}=0
\label{26}
\end{equation}
This is the usual scalar wave equation that appears in the analog gravity with the additional presence of the torsion vector $\vec{K}={\rho}_{0}\vec{v}_{0}{\times}\vec{\Omega}_{0}$. Actually this result was obtained by comparison with the wave equation in the relativistic Riemann-Cartan spacetime. In general relativistic analogue models the fluid equations are expressed in terms of the Riemannian wave equation for a scalar field ${\psi}$ in the form
\begin{equation}
{\Box}^{Riem}{\psi}= 0
\label{27}
\end{equation}
where ${\Box}^{Riem}$ represents the Riemannian D'Lambertian operator given by
\begin{equation}
{\Delta}^{Riem}=\frac{1}{\sqrt{-g}}{\partial}_{i}(\sqrt{-g}g^{ij}{\partial}_{j})
\label{28}
\end{equation}
In this case $(i,j=0,1,2,3)$ where g represents the determinant of the effective Lorentzian metric. Applying minimal coupling of torsion to the metric one obtains the covariant derivative of an arbitrary vector field $B_{k}$ in RC spacetime 
\begin{equation}
{\nabla}_{i}B_{j}={\partial}_{j}B_{j}- {{\Gamma}_{ij}}^{k}B_{k}
\label{29}
\end{equation}
where ${\Gamma}$ is the RC spacetime connection given in terms of the Riemannian-Christoffel connection ${\Gamma}'$ by 
\begin{equation}
{{\Gamma}_{ij}}^{k}={{{\Gamma}'}_{ij}}^{k}-{K_{ij}}^{k}
\label{30}
\end{equation}
where ${K_{ij}}^{k}$ are the components of the contortion tensor. These formulas allow us to write the non-Riemannian D'Lambertian as
\begin{equation}
{\nabla}_{i}{\psi}^{i}={{\nabla}^{Riem}}_{i}{\psi}^{i}+g^{ij}{K_{ij}}^{k}{\psi}_{k}
\label{31}
\end{equation} 
To simplify future computations we consider the trace of contortion as given by $g^{ij}{K_{ij}}^{k}:=K^{k}$. We also define ${\psi}^{i}={\partial}_{i}{\psi}$. This definition allows us to express the non-Riemannian D'Lambertian of a scalar function as 
\begin{equation}
{\Box}{\psi}={{\Box}^{Riem}}{\psi}+ K^{k}{\partial}_{k}{\psi}
\label{32}
\end{equation}
Thus the non-Riemannian wave equation  
\begin{equation}
{\Box}{\psi}=0
\label{33}
\end{equation}
reduces to the following equation
\begin{equation}
{{\Box}^{Riem}}{\psi}= - K^{k}{\partial}_{k}{\psi}
\label{34}
\end{equation}
which justifies the comparison between torsion vector and vorticity in the case of analogue gravity. Actually from the historical point of view is worthwhile to remenber that Elie Cartan started his investigation of torsion by comparison it with torque and to vorticity and translation \cite{1}. In the equation (\ref{26}) the speed $c_{A}$ represents the phase Alfven wave velocity of plasma wave velocity given by
\begin{equation}
{c_{A}}^{2}= \frac{{\mu}^{*}{H_{0}}^{2}}{{\rho}_{0}}
\label{35}
\end{equation}
Thus the effective black hole metric in moving plasma is
\begin{equation}
\sqrt{-g}{g}^{00}= {{\rho}_{0}}({c_{A}}^{2}-{v_{0}}^{2})
\label{36}
\end{equation}
\begin{equation}
\sqrt{-g}{g}^{0j}= {{\rho}_{0}}{{c_{A}}^{2}}({{\vec{v_{0}}}^{T}})_{0j}
\label{37}
\end{equation}
The plasma effective metric is 
\begin{equation}
ds^{2}=\frac{{\rho}_{0}}{c_{A}}[{c_{A}}^{2}dt^{2}-{\delta}_{ij}(dx^{i}-{v_{0}}^{i}dt)(dx^{j}-{v_{0}}^{j}dt)]
\label{38}
\end{equation}
\section{Hawking radiation and horizons in non-Riemannian moving plasmas}
In this section we analyze the issue of the existence of Hawking radiation and horizons in the metric $g_{{\mu}{\nu}}$. From the expression 
\begin{equation} 
g_{00}={c_{A}^{2}-{v_{0}}^{2}}
\label{39}
\end{equation}
we obtain the horizon as the usual black hole condition in general relativity $g_{00}=0$ which in our case yields
\begin{equation} 
{c_{A}}^{2}={v_{h}}^{2}=\frac{{\mu}^{*}{H_{0}}^{2}}{{\rho}_{0}}
\label{40}
\end{equation}
By the definition of Hawking radiation \cite{5}
\begin{equation}
kT_{H}= \frac{\bar{h}}{2{\pi}}\frac{g_{H}}{c_{H}}
\label{41}
\end{equation}
where $g_{H}$ is the surface gravity defined by
\begin{equation}
g_{H}=\frac{1}{2}\frac{d}{dr}[{c_{Pl}}^{2}-{v_{0}}^{2}(r,t)]
\label{42}
\end{equation}
From these expressions we obtain the form of Hawking radiation
\begin{equation}
kT_{H}= \frac{\bar{h}}{2{\pi}}\frac{g_{H}}{\frac{{\mu}^{*}{H_{0}}^{2}}{{\rho}_{0}}}
\label{43}
\end{equation}
This result is similar to the one proposed by De Lorenci et al \cite{4}. For classical fluids it has been proposed by Unruh \cite{5} in 1981 with the purpose of investigating more realistic Hawking effect and sonic spectrum of temperature, which allowed him to proposed the concept of sonic black hole or dumb hole.  
\section{Effective Lorentz symmetry breaking by analogue torsion}
In this section we show that the analogue torsion plays an  important role in the physics of analogue models as it induces the breaking of Lorentz invariance and burefrigence as in the case of the acoustic Lorentz breaking giving by M. Visser \cite{1} in analogy to the same physical phenomena of gravitational spacetime Lorentz violation recently investigated by Kostelecky \cite{12} where he shows that explicitly Lorentz violation is found to be imcompatible with generic RC geometries , but spontaneous Lorentz breaking avoids this problem. Let us consider the wave equation given by the expression  
\begin{equation}
{\rho}_{0}\frac{{\partial}^{2}}{{\partial}t^{2}}{\psi}_{1}+{\rho}_{0}\vec{v}_{0}.{\nabla}\frac{\partial}{{\partial}t}{\psi}_{1}+\vec{K}.{\nabla}{\psi}_{1}=0
\label{44}
\end{equation}
We shall now consider the eikonal approximation in the form
\begin{equation}
{\psi}_{1}= a(x) exp(-i[{\omega}t-\vec{k}.\vec{x}])
\label{45}
\end{equation}
with $a(x)$ a slow varying function of position which is equivalent to ${\nabla}{a(x)}$ being approximately zero throughout the computations. We also do not consider derivatives of the metric. Thus the wave equation with vorticity in the eikonal approximation reads 
\begin{equation}
[{c_{A}}^{2}k^{2}-{\omega}^{2}]-{\beta}{\omega}+i\vec{K}.\vec{k}=0
\label{46}
\end{equation}
\begin{equation}
{\omega}_{\pm}=\frac{-{\beta}\pm{\sqrt{{\beta}^{2}+4i({\theta}-i{c_{A}}^{2}k^{2})}}}{2}
\label{47}
\end{equation}
where to simplify matters we have defined the following quantities: ${\beta}:=\frac{{\vec{v}_{0}}.\vec{k}}{{\rho}_{0}}$ and ${\theta}:=\vec{K}.\vec{k}$. From these quantities and the dispersion formula one notes that the first term is due to the bulk motion of the fluid, the second term is due to the rotational (or analogue torsion) contribution to Lorentz violation while the the last term is the dissipative term. Note that the Lorentz violation is supressed at low momentum. Note also that vorticity adds a term to the bulk matter contribution to ${\omega}_{0}$. Besides we note that the dissipative term does not have the contribution of vorticity.
\section{Conclusions}
We have obtain the effective black hole in moving plasma within the analogue non-Riemannian spacetime where analogue torsion is proportional to the background fluid vorticity similar to what happens in the quantum superfluid case discussed earlier. The Hawking effect is shown to be proportional to the background constant magnetic field squared while the effective Lorentz breaking due to analogue torsion explicitly demonstrated by making use of an eikonal geometric acoustic approximation. The important issue here is that this is analogous to the Riemann-Cartan relativistic case of Lorentz breaking investigated by Kostelecky \cite{11}. 
\section*{Acknowledgements}
\paragraph*{}
I am also very much indebt to Professor W. Unruh, Dr. C. Furtado, Dr. S. Bergliaffa  and Professsor P. S. Letelier for discussions on the subject of this paper. Special thanks go to Professor V. Pagneux for some discussions on acoustic perturbations. Grants from CNPq (Ministry of Science of Brazilian Government) and Universidade do Estado do Rio de Janeiro (UERJ) are gratefully acknowledged.


\begin{thebibliography}{13}
\bibitem{1} For a review see C. Barcelo, S. Liberati and M. Visser, Analogue Gravity in Relativity Living reviews (2005) and M. Novello, M. Visser and G. Volovik, Artificial Black Holes (world scientific) 2002.
\bibitem{2} U. Leonhardt and Pwinicki, Phys. Rev. Lett. 84 (2000) 822. 
\bibitem{3} R. Schuetzhold, G. Plunien and G. Soff, Phys. Rev. Lett 88 (2002),061101.
\bibitem{4} V. De Lorenci, R. Klippert and Yu N. Obukhov, Phys. Rev. D 68 (2003) 061502. V. De Lorenci and R. Klippert, Phys. Rev. D 665 (2002) 064027.
\bibitem{5} W. Unruh, Phys. Rev. Letters, 1351 (1981).
\bibitem{6} L.C. Garcia de Andrade, Phys. Rev. D 70 (2004),64004-1. \bibitem{7} E. Cartan, Riemannian Geometry in an Orthogonal Frame, World Scientific (2001).
\bibitem{8} L.C. Garcia de Andrade, Physics Letters A 339 (2005) 188 and On the necessity of non-Riemannian acoustic spacetime in fluids with vorticity, Phys. Letters A in press.
\bibitem{9} A. C. Eringen and G. A. Maugin,Electrodynamics of Continua II-Fluids and Complex Media (1990), Springer-Verlag. 
\bibitem{10} C. Barcelo, S. Liberati and M. Visser,Towards the observation of Hawking radiation in Bose-Einstein condensates, gr-qc/0110036. M. Visser, Acoustic black holes: horizons,ergosphere, and Hawking radiation,gr-qc/9712010v2. Classical and Quantum Gravity(1998). 
\bibitem{11} V. Alan Kostelecky,Phys. Rev. D 69 (2004)105009.
\bibitem{12} I thank V. Pagueneux for these comments on the acoustics of vortices in water. 
\bibitem{13} D. Kamp, Magnetic Reconnection (2004). 
\end{thebibliography}
\end{document}